\documentclass[floatfix,twocolumn,tighten]{aastex63}
\usepackage{amssymb, amsmath,framed}
\usepackage{textcomp,gensymb}
\usepackage{savesym}
\savesymbol{tablenum}
\usepackage{siunitx}
\restoresymbol{SIX}{tablenum}

\usepackage{bm}
\expandafter\ifx\csname package@font\endcsname\relax\else
 \expandafter\expandafter
 \expandafter\usepackage
 \expandafter\expandafter
 \expandafter{\csname package@font\endcsname}

\fi
\def\bq{\begin{equation}}
\def\eq{\end{equation}}
\def\bqy{\begin{eqnarray}}
\def\eqy{\end{eqnarray}}

\def\p{\partial}

\def\p{\partial}

\def\R{\mathbb{R}}

\begin{document}
\title{\large{Can We Fly to Planet 9?}}

\correspondingauthor{Adam Hibberd}
\email{adam.hibberd@i4is.org}

\correspondingauthor{Manasvi Lingam}
\email{mlingam@fit.edu}

\author{Adam Hibberd}
\affiliation{Initiative for Interstellar Studies (i4is) 27/29 South Lambeth Road London, SW8 1SZ United Kingdom}

\author{Manasvi Lingam}
\affiliation{Department of Aerospace, Physics and Space Sciences, Florida Institute of Technology, Melbourne FL 32901, USA}
\affiliation{Department of Physics and Institute for Fusion Studies, The University of Texas at Austin, Austin, TX 78712, USA}

\author{Andreas M. Hein}
\affiliation{SnT, University of Luxembourg, 29 Avenue J.F Kennedy, L-1855, Luxembourg}
\affiliation{Initiative for Interstellar Studies (i4is) 27/29 South Lambeth Road London, SW8 1SZ United Kingdom}

\begin{abstract}
Planet 9 is an hypothetical object in the outer Solar system, which is as yet undiscovered. It has been speculated that it may be a terrestrial planet or gas/ice giant, or perhaps even a primordial black hole (or dark matter condensate). State-of-the-art models indicate that the semimajor axis of Planet 9 is $\sim 400$ AU. If the location of Planet 9 were to be confirmed and pinpointed in the future, this object constitutes an interesting target for a future space mission to characterize it further. In this paper, we describe various mission architectures for reaching Planet 9 based on a combination of chemical propulsion and flyby maneuvers, as well as more advanced options (with a $\sim 100$ kg spacecraft payload) such as nuclear thermal propulsion (NTP) and laser sails. The ensuing mission duration for solid chemical propellant ranges from 45 years to 75 years, depending on the distance from the Sun for the Solar Oberth maneuver. NTP can achieve flight times of about 40 years with only a Jupiter Oberth maneuver whereas, in contrast, laser sails might engender timescales as little as 7 years. We conclude that Planet 9 is close to the transition point where chemical propulsion approaches its performance limits, and alternative advanced propulsion systems (e.g., NTP and laser sails) apparently become more attractive. \\
\end{abstract}

\section{Introduction} \label{SecIntro}
Ever since the International Astronomical Union (IAU) modified the status of Pluto to that of a ``dwarf planet'' from its initial designation as one of nine planets in the Solar system,\footnote{\url{https://www.iau.org/public/themes/pluto/}} the Solar system is held to comprise eight \emph{known} planets currently. The debate of what exactly constitutes a ``planet'' is still ongoing \citep[e.g.,][]{MGS22}, although many scientists adhere to the taxonomic classification propounded by the IAU. 

After Pluto's status was updated, the hunt for an elusive ``Planet Nine'' far beyond the orbit of Pluto -- alternatively known as ``Planet X'' -- has accordingly intensified, although it must be recognized that this endeavor long predates the aforementioned IAU decision. A historical overview of this field can be found in the recent publication by \citet{BAB19}. The most crucial modern development in this realm arguably concerns the work of \citet{BB16}: the authors theorized that the (ostensibly) improbable orbital clustering of certain Kuiper Belt Objects (KBOs) is strikingly explainable through the existence of a distant planet with mass $\sim 10\,M_\oplus$ on an eccentric orbit. Earlier notable developments in the twenty-first century relating to Planet 9 are reported in \citet{DD14}, \citet{TS14}, and \citet{GSB15}, among other papers.

This noteworthy proposal by \citet{BB16} initiated a plethora of publications centered on myriad aspects of the putative Planet 9. The areas explored in detail include: constraints on the location based on orbital dynamics and/or data from telescopes and spacecraft trajectories \citep{BBat16,DD16,FLMG,GSL16,HP16,MVW16,LI17,ML17,LHPH,CK20,DM20,FDB20,NAB21,BB22,BBB21,MRR22,SN22}, analyses of the clustering of KBOs \citep{BatB16,ST16,MEB17,BB19,DD22}, potential formation mechanisms \citep{BK16,KB16,MRD16}, avenues for detection \citep{CHK16,FML16,GSL16,LM16,JS17,RL19,RL20,AK21,SS22}, impact on dynamical stability and evolution of the Solar system \citep{DDA16,LA16,DV16,BM17,CS21}, effect(s) on solar and planetary obliquities \citep{DL16,BAK17,GDM17,LL22}, and other cognate topics \citep{PBV18,KBA20,LX20,BB21,OT21,NYG22}.

We caution, however, that the exact status and nature of the putative Planet 9 remains unresolved. Some groups have critically reassessed the evidence for the clustering of KBOs \citep{SKB17,BBS20,NGL21,BBS22}, while others have sought to explain the purported clustering through physical mechanisms independent of Planet 9 \citep{ZCT20,ZTC21}. It was even suggested that Planet 9 may be a primordial black hole \citep{SU20}, and methods for detecting this class of objects in the Solar system have been proposed and debated \citep{EW20,HL20,LR20,SL20,MAC22,NHW22}. The possibility that Planet 9 might represent a dark matter condensate of some kind was also advanced \citep{SKK16}.

Although the location and characteristics of Planet 9 -- and even its very existence -- are currently not confirmed, it is nevertheless advantageous to investigate the feasibility of a mission to this object for two major reasons. First, as described hereafter in Section \ref{SciReturn}, launching a mission to Planet 9 would yield a wealth of scientific information about its properties, plausibly more than what can be discerned through observations from ground- and space-based telescopes. Second, irrespective of whether Planet 9 as predicted by \citet{BB16} is real, formulating the specifics of such a mission could serve as a valuable starting point for missions to trans-Neptunian objects (TNOs) \citep{GV21} at similar locations (see also \citealt{MSD11,ZSF21,ZV22}) if promising targets for exploration of this kind are identified.

Therefore, the objective of this paper is to analyze various trajectories and mission designs to the hypothetical Planet 9. To err on the side of caution, we assume that \emph{current} technologies are deployed for the most part, although we comment briefly on the utility of near-future technologies in Sections \ref{SecResults} and \ref{SecDisc}. The orbital parameters for Planet 9 are adopted from \citet{BroB21}, which represents a recent estimate of this object's location. To the best of our knowledge, the only other study that analyzes trajectories to Planet 9 is \citet{CFP22}. Our paper differs from this work in the following major respects: (a) we adopt a conservative estimate for Planet 9's distance to account for the potential necessity of longer mission durations, (b) we do not restrict ourselves to just trajectories but also tackle concrete mission architectures, and (c) we incorporate alternative flyby maneuvers.

\section{Science return and benefits from missions to Planet 9}\label{SciReturn}
Before embarking on summarizing the science benefits that would accrue from sending a mission to Planet 9, it is helpful to draw on an analogy with Pluto. Even though Pluto was discovered in 1930 and much information about this object was garnered from telescope observations, a wealth of new data surpassing the old was yielded by the \emph{New Horizons} flyby of the Pluto–Charon system \citep{SGM18,SMG21}. In particular, the \emph{New Horizons} mission shed light on Pluto's geological diversity, especially its atmosphere, complex surface, and interior \citep{SBE15,GSE16,NMH16}. Additionally, the mission provided insights into Charon and Pluto's small satellites \citep{WBB16}, as well as how this system may have formed.

Note that Pluto has a semimajor axis of $a_P \approx 40$ AU, whereas the corresponding value for Planet 9 is presumed to be nearly ten times higher at $a_9 \sim 380$ AU \citep{BroB21}. The radius of Pluto is $R_P \approx 0.19\,R_\oplus$ while that of Planet 9 may be $R_9 \sim 1.5\,R_\oplus$ (refer to \citealt{BroB21}). Hence, if we posit that both these objects have a similar Bond albedo (which is however not guaranteed to hold true), the ratio of reflected light fluxes ($\delta_R$) from Planet 9 and Pluto at Earth's location can be estimated by using
\begin{equation}\label{deltaR}
   \delta_R \sim \left(\frac{R_9}{R_P}\right)^2 \left(\frac{a_9}{a_P}\right)^{-4} \sim 7.7 \times 10^{-3}. 
\end{equation}
The ratio of thermal emission fluxes ($\delta_E$) from Planet 9 and Pluto at Earth, assuming they possess similar temperatures (see \citealt{LM16}),\footnote{Despite the fact that Planet 9 lies far beyond the orbit of Pluto and might thus be expected to have a lower temperature, its larger size would translate to higher internal heating (thereby raising the temperature), and a non-negligible greenhouse effect cannot be dismissed altogether.} is
\begin{equation}\label{deltaE}
     \delta_E \sim \left(\frac{R_9}{R_P}\right)^2 \left(\frac{a_9}{a_P}\right)^{-2} \sim 0.7.
\end{equation}
Although these calculations are evidently heuristic, they indicate that it is not just harder to detect Planet 9 (compared to Pluto), but also that it may be more challenging to infer its properties from Earth. Therefore, in view of the preceding paragraph (where Pluto was discussed), it does not seem unreasonable to contend that the scientific yield from a flyby mission to Planet 9 -- to say nothing of rendezvous or sample return missions -- could significantly exceed the scientific yield from telescope observations. As per (\ref{deltaR}) and (\ref{deltaE}), the relative gain in scientific return from sending a mission to Planet 9 might even exceed that of Pluto since the former is apparently harder to characterize via remote observations from Earth-based telescopes.

As this work seeks to assess viable mission trajectories and designs, it is not our goal to furnish an exhaustive analysis of the myriad areas that could witness major advances from a mission to Planet 9. With this caveat in mind, we list a few potential avenues below.
\begin{itemize}
    \item As per current estimates, Planet 9 is predicted to have a mass of $M_9 \sim 6\,M_\oplus$ \citep{BroB21}. Planets belonging to this class -- the mass range between Earth and Neptune - are unknown hitherto in the Solar system, but have been detected often in extrasolar systems \citep{FP18}, albeit with typically high irradiances (and temperatures). There is considerable uncertainty surrounding their composition and structure, in particular whether they are endowed with massive H$_2$/He envelopes and/or substantial H$_2$O inventories \citep{JM18,ZJS19,MDA20,VGH20,HDS21,MPC21}. Hence, the characterization of Planet 9 could provide an unparalleled opportunity to investigate planets from this crucial category and acquire in-depth data.
    \item Earth-based observations may conceivably yield information about Planet 9's internal heat budget \citep{CHK16} and its atmospheric composition \citep{FML16}, but obtaining spatially resolved data of its surface (assuming it is well-defined) and its interior seems unlikely. In contrast, the science payload of $\sim 30$ kg of \emph{New Horizons} \citep{WGT08} was sufficient to glean vital insights about the surface and interior of Pluto \citep{SGM18}, like its potential subsurface ocean \citep{NMH16}. With similar payload and flyby distance, a great deal might be inferred about the physical, chemical, and geological processes and composition of of Planet 9 specifically, as well as TNOs (see \citealt{HBS21}).
    \item Many hypotheses have been propounded to explain how Planet 9 formed at its presumed location or migrated there \citep{BK16,KB16,MRD16}. Detailed spacecraft observations could enable us to not only differentiate between these hypotheses but also constrain the formation and dynamical evolution of the outer Solar system. For instance, if instrumentation aboard the spacecraft has the capacity to determine oxygen isotope ratios \citep{NG12,RNC03,IAG20}, they might aid in distinguishing whether Planet 9 was assembled \emph{in situ} in the Solar system or was captured from another star (see \citealt{NG12,RNC03,IAG20}); the latter scenario was theoretically studied by \citet{MRD16}. Likewise, isotopic measurements of hydrogen, nitrogen, carbon, and sulfur -- analogous to those undertaken for comets \citep{JMH09,BCC15} -- are valuable for unraveling the formation history as well as subsequent physical and chemical processes responsible for fractionation on Planet 9. 
    \item Last, but not least, depending on the scientific payload along with the nature of Planet 9's atmosphere and surface (if existent), it may be feasible to survey them for biosignatures. Although the majority of biosignatures have been formulated for Earth-like planets \citep{JLG17,LL21}, there is increasing emphasis on identifying biosignatures that could arise on ``exotic'' worlds \citep{ISM20,LL21}, such as those with anoxic atmospheres dominated by H$_2$ \citep{SBH13,SSR20,MPC21,WSG21,HSP22}, which is a conceivable composition for Planet 9 \citep{FML16}.
\end{itemize}
We underscore, in closing, that we have only described a select few avenues where a mission to Planet 9 is anticipated to substantially advance our knowledge of planetary science. Nonetheless, we hope that this primer serves as an adequate basis for motivating the rest of our technical exposition in the forthcoming sections.

\section{Approach}
The key components of our framework and approach are elucidated in this section. 

\subsection{Orbital Mechanics}
Evidently, in order to perform some kind of meaningful analysis, we must have some notion of Planet 9's position at the present, the velocity and orbit being less relevant. The latter two parameters are of secondary importance because Planet 9's great distance translates to a slow heliocentric speed ($\sim 1.4  \si{.km.s^{-1}}$) and low mean motion ($\sim 0.04  \si{.\degree. yr^{-1}}$). Consequently, as the orbit of Planet 9 would only be required for extrapolating forwards the trajectory, such an extrapolation would yield dubious benefits, especially as the estimates of these parameters are subject to large uncertainties.

\begin{table}
\label{table:Planet9}
\caption{Ephemeris and orbital elements adopted for calculating Planet 9 missions}
\hspace{-0.8cm}
\begin{tabular}{|c|c|}

\hline
\textbf{Ephemeris}                                                                    &             \\ \hline
Sun Distance ($\si{AU}$)                                                            & 450         \\ \hline
RA ($\si{\degree}$)                                                                     & 65          \\ \hline
DEC ($\si{\degree}$)                                                                    & 20          \\ \hline
\begin{tabular}[c]{@{}c@{}}Mean Motion \\ ($\si{\degree .yr^{-1}}$)\end{tabular}            & 0.04        \\ \hline
\textbf{Orbital Parameters}                                                           &  Relative to Ecliptic           \\ \hline
Perihelion ($\si{AU}$)                                                              & 450         \\ \hline
Eccentricity                                                                 & 0.0         \\ \hline
\begin{tabular}[c]{@{}c@{}}Argument of \\ Perihelion ($\si{\degree}$)\end{tabular}      & 246         \\ \hline
\begin{tabular}[c]{@{}c@{}}Longitude of \\ Ascending Node ($\si{\degree}$)\end{tabular} & 180         \\ \hline
Inclination ($\si{\degree}$)                                                            & 1.43        \\ \hline
Epoch of Perihelion ($\si{\degree}$)                                                          & 2035 JAN 01 \\ \hline
\end{tabular}
\end{table}

To reinforce the prior argument, we comment briefly on Planet 9's change of position over the duration of the spacecraft's full journey. As demonstrated hereafter, we end up with overall flight times of around $60$ years or so for chemical propulsion, which from Table \ref{table:Planet9} we can calculate will result in a change of longitude of about $2.4 \si{\degree}$, well within the range of longitudinal uncertainty of Planet 9 as per current theoretical and empirical constraints. Nevertheless for the purposes of the research conducted here, some fiducial orbital parameters are required, in addition to Planet 9's position. Values adopted for the position and velocity of Planet 9, as well as orbital elements are provided in Table \ref{table:Planet9}. 

For the analysis here, save for one exception, the earliest launch date for a mission is specified as 2030 JAN 01 and the latest as 2043 JAN 01, in order to permit ample opportunity for an entire Jovian cycle of 11.9 years to complete. Given that Jupiter completes about a twelfth of a cycle per year (approximately $30 \si{\degree}$), for an opportunity arising in a particular year, Jupiter may be displaced from optimum by up to $30 \si{\degree}$. For this reason it may occasionally be of interest to compare the results of a ``true'' Jupiter location within the range mentioned above against a ``theoretically optimal'' Jupiter position, were Jupiter able to occupy any location in its orbit irrespective of its positional dependency on time.

Note that the launch range above, 2030-2043, aligns pretty well with the timeline elaborated by the \emph{Interstellar Probe} concept report \citep{9438249,MWG22}. The above time interval implicitly assumes that astronomers have discovered Planet 9 by 2030 and have accurately determined its orbital elements, thus allowing at least 8 years for telescopes to (1) detect the planet and (2) conduct observations to determine the orbit of the planet.

\subsection{Mission Trajectories}
The trajectory studies here were conducted using Optimum Interplanetary Trajectory Software (OITS); refer to \citet{AH22}. At its core is an algorithm which solves the Lambert problem using the Universal Variable Formulation as elaborated in \cite{Bate1971}. Ignoring multiple orbital cycles, there are two solutions to this problem, ``short way'' (sw) and ``long way'' (lw). These constitute two different orbits, with different orbital parameters, but with the same plane, defined by the two position vectors. Thus if the change in true-anomaly along the short way is designated $\theta$ where $0 \leq \theta \leq \pi$, then the long way corresponds to $2\pi - \theta$. By exploiting the NASA SPICE toolkit,\footnote{\url{https://naif.jpl.nasa.gov/naif/toolkit.html}} in conjunction with the selection of appropriate ephemeris data available from the NASA Horizons service (in the form of binary SPICE kernel files), extremely accurate ephemerides of a particular object as a function of time can be determined. 

When dealing with more than 2 celestial bodies, amounting to $n > 2$ bodies, the Lambert problem can be solved for each consecutive pair of celestial bodies along the trajectory. This method leads to a possible $2^{n-1}$ permutations of interplanetary trajectory, comprising long ways and short ways. For each non-terminal encounter, there are four possible encounter trajectories with respect to the celestial body, an arrival pair of hyperbolic excess velocities (corresponding to sw \& lw of the preceding interplanetary trajectory) and a departure pair (sw \& lw for the subsequent one). Connecting hyperbolae with respect to the body are then computed for each arrival and departure combination, assuming a single $\Delta V$ is applied at the periapsis point, aligned with the plane defined by these hyperbolic excess velocities and tangential to the trajectory.  

Two Non-Linear Programming (NLP) solvers were used for this study, namely, NOMAD and MIDACO. 
To reach Planet 9 or alternatively a Sednoid, three distinct trajectory strategies are considered here, shown in Table \ref{table:Trajectory}. For OITS, a Solar Oberth maneuver (SOM) can be modeled as an ``Intermediate Point'' \citep{AH22} where the heliocentric radial distance is specified, but the heliocentric longitude and latitude are additional optimization parameters for OITS. This framework can also model a Deep Space Maneuver (DSM).

\begin{table*}[]
\label{table:Trajectory}
\caption{Trajectory Options for Planet 9}
\begin{tabular}{|c|c|c|c|}
\hline
  & \textbf{Trajectory Option}        & \textbf{Trajectory Description}                        & \textbf{Abbreviation} \\ \hline
1 & Passive Jupiter flyby    & Pure Jupiter GA without thrust                & PJGA         \\ \hline
2 & Powered flyby of Jupiter & Combined Jupiter GA with Jupiter Oberth maneuver - JOM & JOM          \\ \hline
3 & Powered flyby of the sun & Solar Oberth maneuver - SOM                            & SOM          \\ \hline
\end{tabular}
\end{table*}

Furthermore, depending on the context, two different optimization criteria (i.e., the so-called ``objective functions'') can then be applied in OITS -- either minimizing $\Delta V$ or flight duration. With regards to the former, a constraint is required for the overall flight duration, and with respect to the latter, OITS allows for a separate $\Delta V$ constraint to be specified at each encounter.

The terminology adopted herein for abbreviating the different mission trajectory scenarios departs slightly from tradition as both home planet (Earth) and target planet (P9) are included in the sequence. Thus E-J-P9 refers to a trajectory from Earth to Jupiter to Planet 9. This extension of the current standard is so that the abbreviation more accurately reflects the precise inputs required by OITS to generate the trajectory. This trajectory, E-J-P9, is an example of a Jupiter Oberth maneuver (JOM) to Planet 9. As a further example, in the case of a SOM we have E-J-3SR-P9, where 3SR indicates the perihelion distance of the SOM from the center of the Sun -- to wit, $3$ Solar radii ($3\,R_\odot$); note that $1\,R_\odot$ (1SR) is equivalent to 0.00465 $\si{AU}$.

A note on $V_{\infty}$ Leveraging Maneuvers (VLMs) is warranted. These maneuvers represent a mechanism by which the $\Delta V$ necessary to travel to Jupiter could be reduced. In addition, VLMs permit longer duration launch windows as well as a lower magnitude $C_{3}$ at launch, although the latter can be offset to an extent by the increased $\Delta V$ required at the Earth return. They involve a launcher injection into a heliocentric elliptical orbit with a DSM at aphelion. A return to Earth then ensues whereupon a powered gravity assist (GA) is executed to transfer to Jupiter. There are two aphelia distances investigated in this study, 2.2 \si{AU} and 3.2 \si{AU}, indicating a resonance of 2 years or 3 years respectively.

The minimum periapsis altitude at each planetary encounter is chosen as 200 \si{km}, thus implying that the NLP will seek trajectories for which the $\Delta V$ application occurs at altitudes greater than this. 

\subsection{Heat Shield Calculations}\label{SSecHeatS}
An equation is required here that would relate the heat shield mass, $M_{hs}$, to:
\begin{enumerate}
    \item total mass of the spacecraft, $M_{sc}$
    \item solar distance $R_{sc}$ of the spacecraft at perihelion 
\end{enumerate}
To this end, scaling of the relevant variables can be performed by benchmarking against an already-known, tried and tested reference for which data is readily available such as the well-known Parker Solar Probe (PSP) for example \citep{DEG19}.

Let us denote the mass of the PSP as $M_{psp}$ and its closest approach to the sun as $R_{psp}$. It is reasonable to suppose that the mass of the heat shield for a spacecraft, $M_{hs}$, must be proportional to the surface area of the spacecraft (as it must encompass a fixed fraction of the spacecraft), so therefore we end up with
\begin{equation}\label{MassScalv1}
M_{hs} \propto \left( \frac{M_{sc}}{M_{psp}} \right)^{2/3}.  
\end{equation} 
Thus, by means of this heuristic scaling, we have partially accomplished item 1, with the rest of the procedure described hereafter. The exponent of $2/3$ appearing in the RHS of the above equation is a consequence of the area-volume scaling and the linear volume-mass relationship for fixed mass density.

For tackling item 2, let us now denote the temperature of the heat shield exposed to the sun as $T$ and the solar flux as $S$, the Stefan-Boltzmann law implies
\begin{equation}
    T \propto S^{1/4}.
\end{equation}
Furthermore, the solar flux $S$ falls off with the inverse square of the distance as $R_{sc}^{-2}$, so we arrive at
\begin{equation}\label{TRrel}
    T \propto R_{sc}^{-1/2}.
\end{equation}
In addition, presuming that heat transfer through the shield is via conduction, then for the specified heat shield thickness $X_{hs}$, it follows that
\begin{equation}
    X_{hs} \propto T,
\end{equation}
for fixed thermal conductivity. Combining the above equation with (\ref{TRrel}) yields
\begin{equation}\label{ThickShield}
    X_{hs} \propto  R_{sc}^{-1/2}.
\end{equation}
For given surface area, the mass of heat shield is linearly proportional to thickness $X_{hs}$, which leads us to
\begin{equation}\label{MassScalv2}
M_{hs} \propto X_{hs}
\end{equation}
Thus, on combining (\ref{MassScalv1}), (\ref{ThickShield}), and (\ref{MassScalv2}), we finally obtain the relationship for the shield mass,
\begin{equation}\label{MassScal}
M_{hs} = M_{psp,hs} \left(\frac{R_{psp}}{R_{sc}} \right)^{1/2} \left(\frac{M_{sc}}{M_{psp}} \right)^{2/3},    
\end{equation}
where $M_{psp,hs}$ is the PSP's heat shield mass, completing the implementation of item 1 delineated previously. 

\subsection{Payload Masses for Chemical Propulsion}
To determine the corresponding payload masses, five parameters are required in total:
\begin{enumerate}
    \item launch vehicle used
    \item the 'Characteristic Energy', $C_{3}$ at launch
    \item the in-flight $\Delta V$ required leading up to the Oberth maneuver 
    \item the $\Delta V$ required at the Oberth maneuver (which might be either JOM or SOM)
    \item the performance of the propulsion systems needed to generate 3 \& 4 above    \end{enumerate}
As far as (1) \& (2) are concerned, the only launch vehicle capable of delivering a sufficiently massive payload to the Earth escape orbits studied here (thence allowing the installation of two rocket stages for the spacecraft's Oberth maneuver), and for which data is available, will be NASA's Space Launch System (SLS) Block 2 variant.\footnote{\url{https://www.nasa.gov/exploration/systems/sls/fs/sls.html}} The calculations can be readily updated if and when data for viable alternatives becomes available. 

Generally, where the trajectory option studied requires the instantiation of item (3) above, a dedicated restartable stage with a hypergolic liquid propellant combination of $MMH$ and $N_{2}O_{4}$ is assumed for all these in-flight maneuvers, with $I_{sp} = 341 \si{s}$ and ratio of dry-mass to wet-mass of $p = 0.1$. With regard to (4) \& (5), either one solid propellant stage from the choice of STAR 75, STAR 63F or STAR 48B is selected, or if there is sufficient spare capacity, two stages comprising a pair of these boosters is allocated. The precise combination depends on the mass available after accounting for items (1)-(3), as well as the required magnitude of (4) necessary for the Oberth maneuver.

Finally it should be emphasized that only flyby missions of Planet 9 shall be considered here. A rendezvous mission, where the spacecraft applies an extra thrust to match velocities with Planet 9, would evidently require more $\Delta V$, which due to Planet 9's low speed would be roughly equal to the spacecraft's heliocentric hyperbolic excess speed. Where the objective function is to minimize flight duration, conducting a rendezvous would have little consequence on the optimal trajectories solved by OITS. Where the purpose is to minimize $\Delta V$, then the effect would be to increase the optimal $\Delta V$ achieved by OITS by a magnitude nearly equal to the spacecraft's heliocentric hyperbolic excess.
\begin{figure}[ht]
\hspace*{-1.0cm}
\includegraphics[scale=0.36]{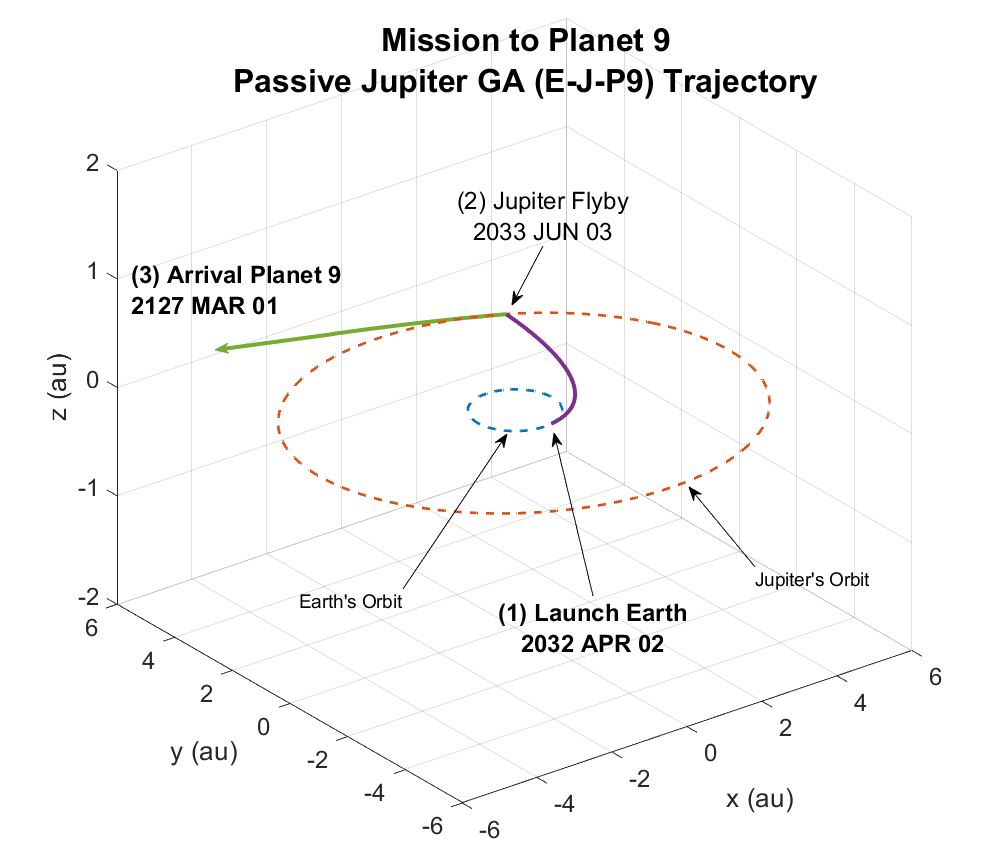}
\caption{Trajectory, E-J-P9 Passive Jupiter GA (without JOM)}
\label{fig:Passive2}
\end{figure}
\begin{figure}[ht]
\hspace*{-1.0cm}
\includegraphics[scale=0.30]{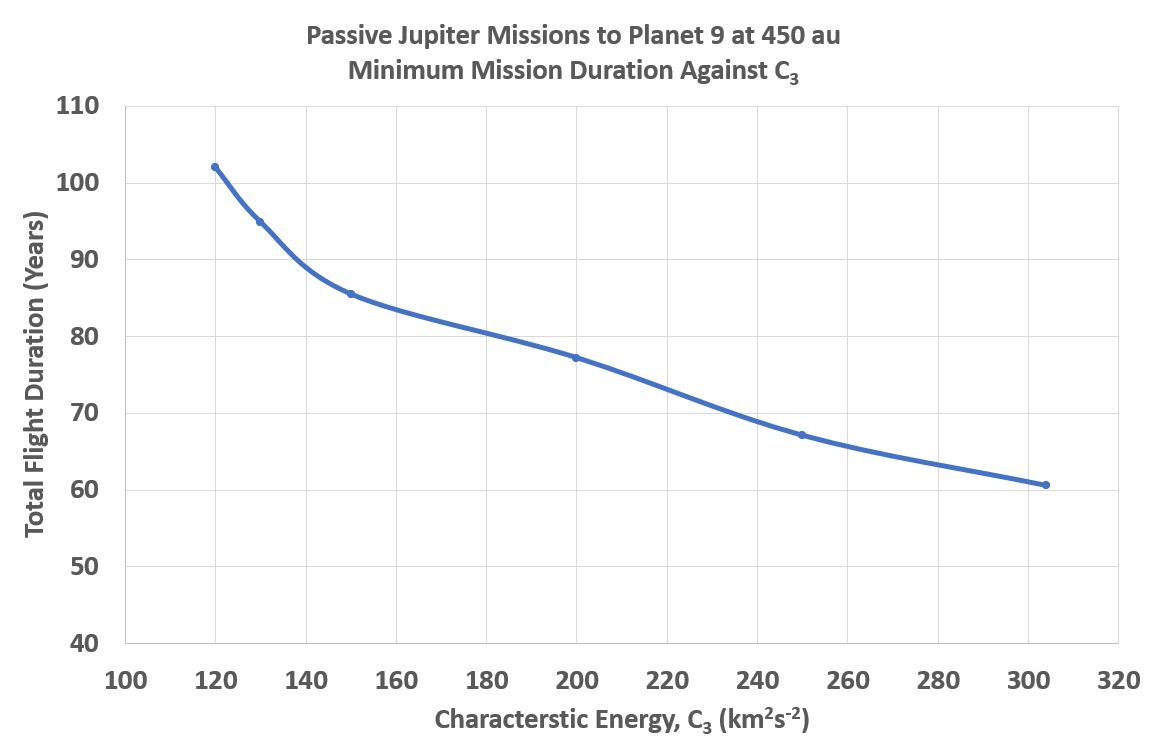}
\caption{Trajectory E-J-P9 Passive Jupiter GA (PJGA), Flight Duration against launch $C_{3}$}
\label{fig:Passive1}
\end{figure}

\section{Results}\label{SecResults}
In this section, we describe the chief results for solid chemical propellant as well as some alternatives.

\subsection{Passive Jupiter GA (PJGA)}
The objective function specified here is flight duration. An example PJGA trajectory is provided in Figure \ref{fig:Passive2}. As in the \emph{Interstellar Probe} concept report \citep{9438249}, a SLS Block 2 is utilized to deliver the spacecraft to Jupiter. In \citep{9438249}, a combined ATLAS V Centaur third stage and STAR 48B fourth stage is deployed as their baseline mission to leverage a $C_{3} = 304.07 \si{.km^{2}.s^{-2}}$, with an eventual payload mass to Jupiter of $\sim 860 \si{.kg}$. Figure \ref{fig:Passive1} depicts total flight duration to Planet 9 in years against launch $C_{3}$. Observe that the flight duration is just over 60 years when the \emph{Interstellar Probe} baseline mission parameters are employed. Table \ref{table:Passive_Data} provides the numerical data for this trajectory with an arrival speed at Planet 9 of $35.4 \,\si{km.s^{-1}}$, which amounts to $\sim 7.47\, \si{AU/yr}$ at $450 \si{AU}$.

\subsection{JOM \& SOM}

\begin{figure}[ht]
\hspace*{-1.0cm}
\includegraphics[scale=0.33]{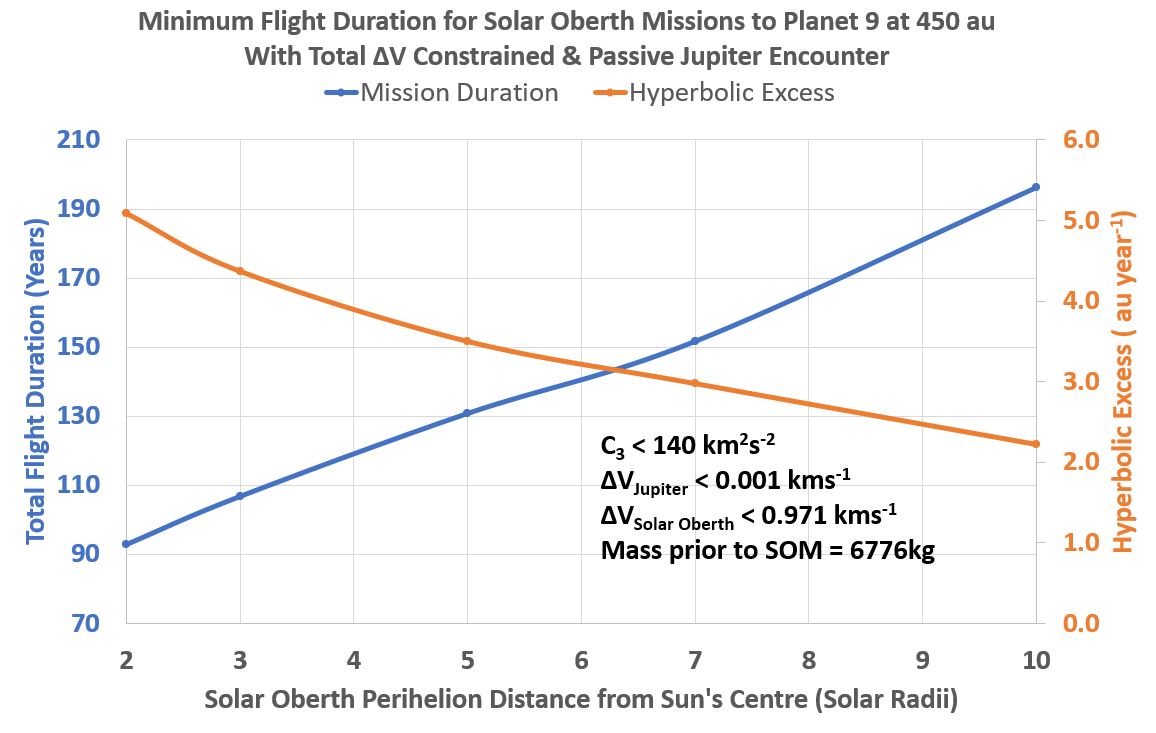}
\caption{Trajectory E-J-SOM-P9 Passive Jupiter GA, Flight Duration and Heliocentric Hyperbolic against SOM perihelion}
\label{fig:Pass_SOM}
\end{figure}

\begin{figure}[ht]
\hspace*{-1.0cm}
\includegraphics[scale=0.33]{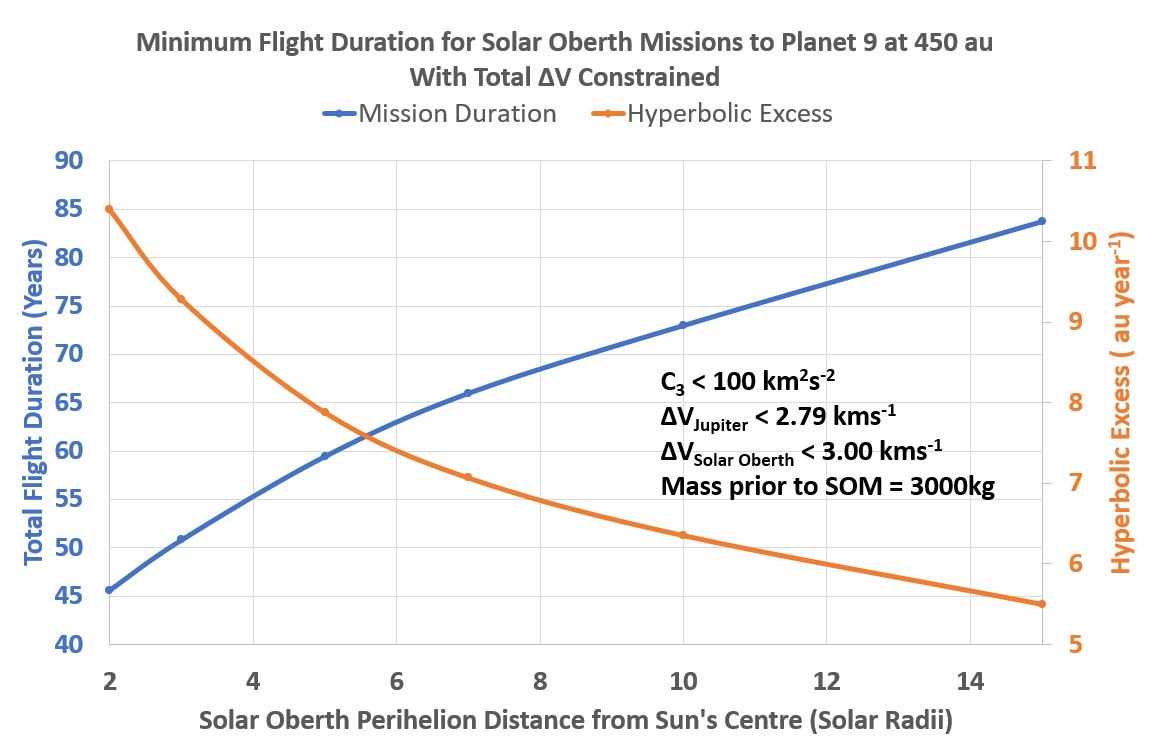}
\caption{Trajectory E-J-SOM-P9 with $\Delta V$ at Jupiter, Flight Duration and Heliocentric Hyperbolic against SOM perihelion}
\label{fig:Pow_SOM}
\end{figure}

The results of the SOM analysis are presented first. In this case, the objective function applied is flight duration. Two distinct scenarios are considered:
\begin{enumerate}
    \item Figure \ref{fig:Pass_SOM} depicts optimal flight duration and heliocentric hyperbolic excess speed adopting a SOM with $\Delta V < 0.97 \si{.km.s^{-1}}$ preceded by a passive Jupiter flyby.
    \item Figure \ref{fig:Pow_SOM} provides the same parameters for a SOM with $\Delta V < 3.0 \si{.km.s^{-1}}$ preceded by a powered Jupiter flyby with $\Delta V$ at Jupiter of $< 2.79 \si{.km.s^{-1}}$.
\end{enumerate}

The $C_{3}$ and $\Delta V$ values for the first scenario stated above are chosen to match those in Figure D-2 of \cite{9438249} which reaches a SOM perihelion of 3SR. Note that in Figure \ref{fig:Pass_SOM}, the hyperbolic excess at 3SR is 4.4 $\si{AU/yr}$, lower than that outlined in Figure D-2 of approx 4.8 $\si{AU/yr}$. This discrepancy is most likely a consequence of the \emph{Interstellar Probe} report \citep{9438249} targeting an optimal solar latitude, whereas here we assume a latitude of Planet 9 of roughly -1.4$\si{\degree}$. For the 3SR case, the payload mass to Planet 9 -- invoked for generating Figure \ref{fig:Pass_SOM} -- is taken from the \emph{Interstellar Probe} report and equals 900 $\si{kg}$, after deducting the heat shield mass (653 $\si{kg}$). As the perihelion distance increases, the solar flux reduces accordingly, indicating that for perihelia $>$ 3SR, the spacecraft mass to Planet 9 will exceed 900 $\si{kg}$ (see Section \ref{SSecHeatS}).

Case 2 introduced above, namely Figure \ref{fig:Pow_SOM}, illustrates the benefit of delivering a higher kick at the SOM, i.e. 3.0 km/s. In order to leverage such a kick, we need a higher spacecraft mass at Jupiter which ultimately necessitates a lower launch $C_{3}$, in this case 100 $\si{km^{2}.s^{-2}}$ delivering a mass of 9000 $\si{kg}$ to Jupiter (utilizing a SLS Block 2 with a Centaur D upper stage). A dedicated liquid propellant stage is exploited to conduct the burn at perijove of $\Delta V = 2.79  \si{.km.s^{-1}}$ leaving a mass of 3000 $\si{kg}$ for the SOM. A STAR 49B can subsequently provide the kick of 3.0 km/s which leaves for the 3SR case a mass of $\sim 520 \si{kg}$ after the heat shield is deducted using (\ref{MassScal}). Larger masses are achievable for higher perihelia, but with commensurately longer mission durations.

\begin{figure*}[ht]
\centering
\includegraphics[scale=0.535]{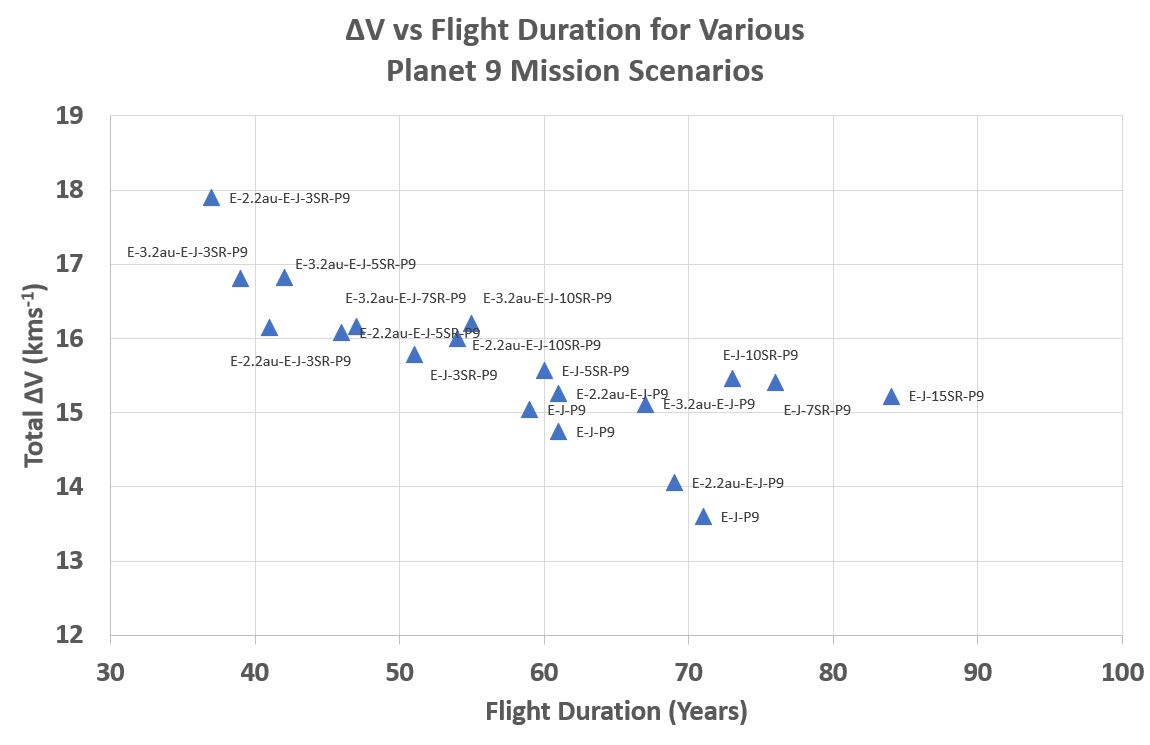}
\caption{Scatter plot of $\Delta V$ against Flight Duration for all JOM \& SOM Missions Studied}
\label{fig:DV_JOMSOM}
\end{figure*}

\begin{figure*}[ht]
\centering
\includegraphics[scale=0.535]{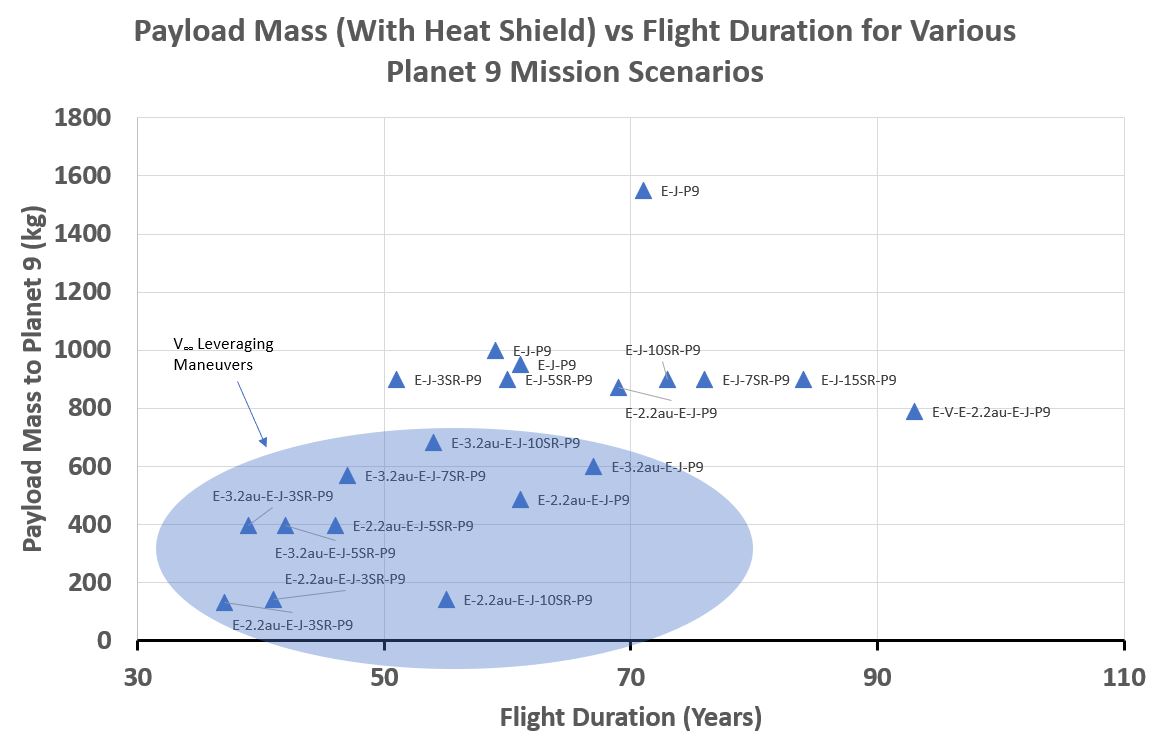}
\caption{Scatter plot of Total Payload Mass against Flight Duration for all JOM \& SOM Missions Studied}
\label{fig:PM_JOMSOM}
\end{figure*}

\begin{figure*}[ht]
\centering
\includegraphics[scale=0.535]{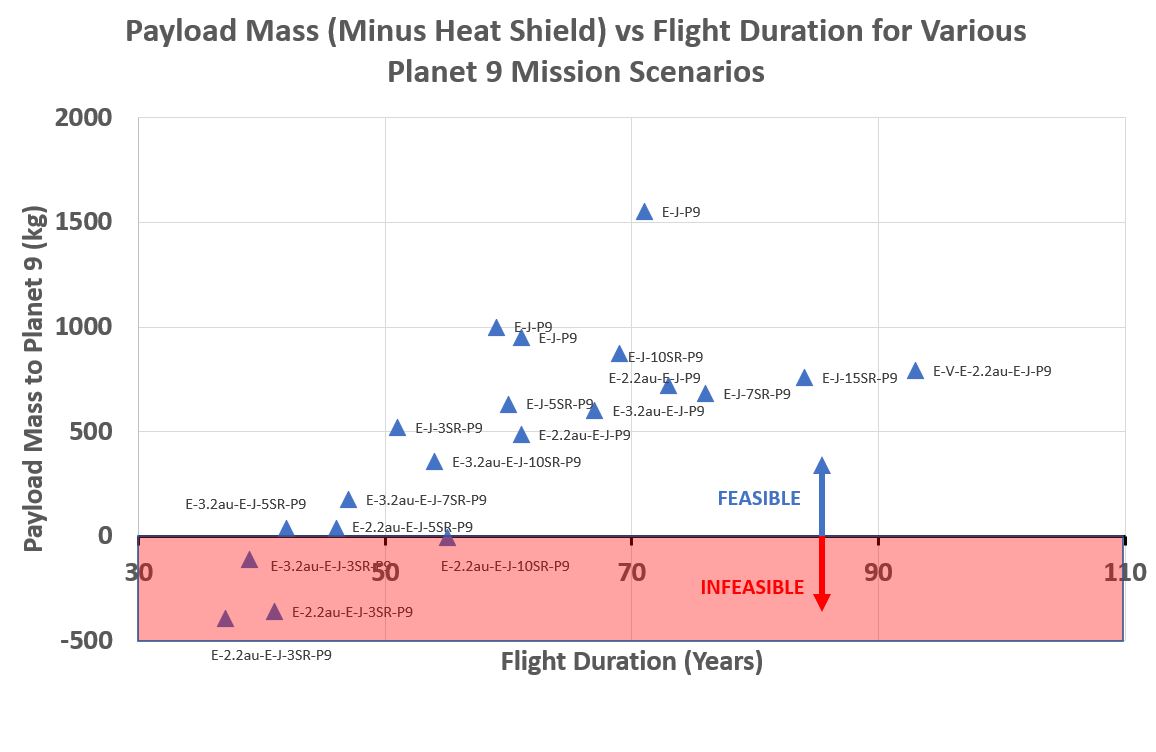}
\caption{Scatter plot of Payload Mass without Heat Shield against Flight Duration for all JOM \& SOM Missions Studied}
\label{fig:HS_JOMSOM}
\end{figure*}

A summary of all mission scenarios, JOM \& SOM is outlined in Figures \ref{fig:DV_JOMSOM}, \ref{fig:PM_JOMSOM} \& \ref{fig:HS_JOMSOM}. Referring to Figure \ref{fig:PM_JOMSOM}, observe that the fastest trajectories to Planet 9 are SOM scenarios that exploit a leveraging maneuver of some kind, though the corresponding price is a lower payload mass. The overall fastest trajectory to Planet 9 is E-2.2-E-J-3SR-P9 yielding arrival after only 37 years. However it also has the lowest total payload mass of 133 $\si{kg}$ and when the required heat shield mass is deducted (consult Figure \ref{fig:HS_JOMSOM}) then we find this mission is actually rendered infeasible using PSP heat shield technology. This result emphasizes the importance of comparing like with like when gauging JOM \& SOM trajectories. We shall thus reference Figure \ref{fig:HS_JOMSOM} hereon. 

As seen from this figure, ignoring those missions which are dangerously close to negative payload masses, the best performance with respect to flight duration is E-3.2-E-J-7SR-P9, requiring 47 years, which is a full decade longer than the mission we excluded above. After this comes the trajectory E-J-3SR-P9 -- which is not the same case as that elucidated in Figure D-2 of the \emph{Interstellar Probe} report \citep{9438249}, as it corresponds to the alternative $C_{3}$ and $\Delta V$ allocation outlined in Figure \ref{fig:Pow_SOM} and delineated previously -- with a flight duration of approximately 50 years. For comparison, the SOM equivalent to \citet[Figure D-2]{9438249} is provided in Figure \ref{fig:Pass_SOM} and would take over 100 years. 

\begin{table*}[]
\caption{Numerical Data for E-J-P9 (PJGA) with $C_{3} = 304.07 \si{.km^{2}.s^{-2}}$}
\label{table:Passive_Data}
\centering
\begin{tabular}{|c|c|c|c|}
\hline
          & 1           & 2           & 3                \\ \hline
Encounter & Earth       & Jupiter     & Planet 9 @ 450 \si{AU} \\ \hline
Time      & 2033 MAY 20 & 2034 FEB 18 & 2093 DEC 01      \\ \hline
\begin{tabular}[c]{@{}c@{}}Arrival Speed \\ (\si{km.s^{-1}})\end{tabular}     & 0.0000  & 27.1881 & 35.4164 \\ \hline
\begin{tabular}[c]{@{}c@{}}Departure Speed \\ (\si{km.s^{-1}})\end{tabular}   & 17.4376 & 27.1902 & 35.4164 \\ \hline
\begin{tabular}[c]{@{}c@{}}DeltaV \\ (\si{km.s^{-1}})\end{tabular}            & 17.4376 & 0.0010  & 0.0000  \\ \hline
\begin{tabular}[c]{@{}c@{}}Cumulative DeltaV \\ (\si{km.s^{-1}})\end{tabular} & 17.4376 & 17.4386 & 17.4386 \\ \hline
\begin{tabular}[c]{@{}c@{}}Altitude Periapsis \\ (\si{km})\end{tabular}  & N/A     & 34110.2 & N/A     \\ \hline
\end{tabular}
\end{table*}

\subsection{Using \texorpdfstring{$LH_2 \,\&\, LOX$}\space{} Propellants}\label{SSecCryo}
With a higher specific impulse than solid propellant ($I_{sp} = 451 \si{s}$ as opposed to $\sim 300 \si{s}$ for solid), the combination of $LH_{2}/LOX$ liquid cryogens evinces the potential to dramatically enhance the performance of interplanetary missions. The key issue is one of achieving sufficiently low temperatures for long-term storage in space where the solar environment can lead to significant heating and boil-off, particularly of $LH_{2}$. One potential solution requiring no mass-costly on-board cryocoolers is to subcool the $LH_{2}$ isobarically to temperatures of less than $16\, \si{K}$ whilst the spacecraft is installed on the launcher using compact ground support equipment \citep{Mustafi2009}. An outstanding issue insofar as this technology is concerned is achieving sufficient compactness to enable the required level of refrigeration to be achieved on the launch pad.

In the spirit of the analysis conducted in \cite{MDF16,CPTOPS}, we shall assume the required level of on-ground subcooling can indeed be performed to satisfaction and therefore as a first order estimate, we assume that no onboard cryocooler mass is necessary. \cite{CPTOPS} employed an ATLAS V AV551 launcher. However, when working with the launch timescales of the missions proposed here, we can suppose that the SLS Block 2 will be available, thereupon enabling a wholesale upscale of the spacecraft dimensions.

When this upscaling is taken into consideration, we estimate a total propellant mass of 11077 \si{kg} and a dry mass of 2083 \si{kg} together with a combined spacecraft payload mass of 100 \si{kg} (which includes science payload, high-gain antenna (HGA), and so forth); this mass breakdown translates to an available $\Delta V$ of 7.97 \si{km.s^{-1}}. Let us suppose we invoke a JOM scenario for this mission and this $\Delta V$ kick is exclusively applied at Jupiter. It should be noted that an SLS Block 2 with Centaur third stage can deliver the aforementioned total mass of 13260 \si{kg} to Jupiter with a $C_3 = 94\, \si{km^2.s^{-2}}$.

Applying OITS with an optimally placed Jupiter encounter and employing the above parameters, we find an arrival at Planet 9 approximately 51 years after launch. This performance level is comparable to the SOMs using solid propellant and summarized in Figure \ref{fig:HS_JOMSOM}, and represents a notable improvement on the alternative E-J-P9 options using solid propellant, which required flight durations of $\sim 60$ years.  

\subsection{Using NTP with \texorpdfstring{$LH_{2}$}\space{}  Propellant}
Nuclear Thermal Propulsion (NTP), as the name indicates, involves fission of uranium isotopes (typically $^{235}U$) wherein the energy released by this fission into smaller isotopes is utilizable in various ways. One common method (NTP in essence) is to heat a cryogenic propellant like $LH_2$ (which also acts as a coolant). This $LH_2$ is then expelled from an engine nozzle with high exhaust velocity, thus giving rise to the nuclear thermal rocket. Using $LH_2$ as propellant has the added benefit of a low molecular mass because specific impulse is proportional to $1/\sqrt{\mathrm{Molar \, Mass}}$, indicating that its use is near-optimal for NTP systems. Specific impulses of at least twice that attainable by chemical rockets can be achieved in principle \citep{GH15}

As a consequence of extensive testing of NTP by the US government sponsored Rover and NERVA programs from 1955 to 1972 \citep{WALTON1991}, NTP has a surprisingly high technology readiness level (TRL) of 5-6, even though there have been no in-flight NTP trials as yet; safety and the possibility of contamination are two major reason for this absence. Despite this caveat, for the sake of comparison, and also taking advantage of the considerable literature on the subject of different types of NTP, we shall adopt and investigate a fiducial NTP system for a mission to Planet 9.
  
The type of NTP assumed is that of \cite{YOUINOU2022104237}, which attempts to derive a design in general conformance with previous Mars DRAs (Design Reference Architectures) \citep{5446736}. The NTP thrust is 66.7 \si{kN} with $I_{sp} \sim 910\si{s}$. We use the highest thrust to weight ratio derived in this study of 10.9, thus translating to an engine mass of $624 \si{kg}$. We assume the same subcooled $LH_2$ adopted earlier for the $LH_2$ \& $LOX$ analysis (this time without any LOX - all the propellant mass is $LH_2$), obviating the need for heavy Brayton-cycle cryocoolers. We further assume the same 100 \si{kg} spacecraft payload (including the HGA) and an SLS with Centaur 3rd stage; the trajectory adopted is that of a JOM with optimally placed Jupiter, with the $\Delta V$ available at Jupiter being 14.386 km/s.
  
Through the use of OITS, we determine that the resulting mission duration for this study is 41 years, a ten year improvement over the cryogenic $LH_2$ \& $LOX$ chemical option outlined in Section \ref{SSecCryo}.

\subsection{Laser Sails}
The notion of employing radiation pressure for propellant-free propulsion (i.e., light sails), derived either from sunlight or laser arrays, is nearly a century old in its modern form \citep{LL21}. More specifically, laser-driven light sails have sprung to the forefront in recent times owing to multiple proposals.

The \emph{Breakthrough Starshot Initiative} \citep{PD17,Parkin_2018},\footnote{\url{https://breakthroughinitiatives.org/initiative/3}} inaugurated in 2016, intends to use powerful Earth-based lasers beamed towards miniature spacecraft of mass $\sim 1$ g, thereby achieving rapid acceleration to speeds that are a sizable fraction of the speed of light. Likewise, \cite{HEIN2019552} outlined the possibility of reaching the interstellar object 1I/'Oumuamua by employing this type of propulsion; detailed analyses of laser sail missions to 1I/'Oumuamua are furnished in \cite{hibberd2020project} by using a speed of 300 km/s. Other recent proposals have suggested that subrelativistic laser sail technology may be exploited to access objects situated far-away in our solar system. 

For instance, \citet{TML20} delineated a light sail architecture to attain a velocity of $0.001$c (i.e., 300 km/s) by utilizing laser arrays with a total power of 3 to 29 GW for spacecraft ranging from 1 $\si{kg}$ to 100 $\si{kg}$, respectively. \citet[Table 2]{KP22} formulated a comprehensive cost-optimal model and determined a 10 kg payload can be accelerated to 300 km/s with peak power of 2.5 GW, capital expenditure of $\$610$ M, and operational expenditure of $\$58$ M per mission.

In fact, \citet{EW20} suggested that laser sails could be harnessed to probe the nature of Planet 9 (viz., ascertain whether it is a primordial black hole) by measuring how its gravitational field affects the spacecraft's trajectory; see also \citet{CL17} and \citet{LR20} for related analyses. \citet{HL20} highlighted the complicating effects of drag and electromagnetic forces on such measurements if the laser sails are relativistic. However, if the location of Planet 9 were to be accurately determined through some other method, subrelativistic laser sails may be deployed (in isolation or en masse) to characterize Planet 9 \citep{TML20,HAH20}.

By implementing OITS with a departure velocity of 300 km/s, it is found that Planet 9 could be reached by laser-sail in a short span of 6.5-7 years, a dramatic improvement over the previous options elaborated in this paper. The downside, of course, is that the TRL of such laser-driven light sails is low because most proposals, including those elucidated in the preceding paragraphs, remain purely conceptual as of now.

\section{Discussion}\label{SecDisc}
The results, upon assuming canonical solid propellants, indicate that trip times for a SOM (preceded by a Jupiter encounter) start with 45 years for perihelion distance of 2 Solar radii and increases to about 70-80 years for 10 Solar radii, a distance similar to the one associated with the Parker Solar Probe. Such trip times are considerably higher than those for existing space missions to survey planets in the outer Solar system. The Voyager probes have now exceeded 40 years of mission duration, but they surveyed the outer Solar system planets (i.e., their targets) in a span of $\sim 10$ years after launch.\footnote{Proposed missions to KBOs, which are much closer than Planet 9 to the Sun, accordingly necessitate shorter flight times of $\sim 25$ years \citep{ZFA19}.} Due to programmatic reasons, it might be unlikely that a mission to Planet 9 based on solid chemical propulsion would receive funding if missions predicated on alternative propulsion technologies were to promise a quicker and more substantial science return.

Bearing this last point in mind, there are three alternative mission designs analyzed in this paper: chemical cryogenic propellants $LH_2$ \& $LOX$, NTP using propellant $LH_2$, and laser sail accelerated by a 29 $\si{GW}$ array to $0.1\%$ the speed of light. All three alternatives commonly assume a 100 \si{kg} spacecraft payload is sent to Planet 9; in comparison, the \emph{New Horizons} spacecraft possessed a total launch mass of $\sim 500$ \si{kg} and $\sim 30$ \si{kg} science payload \citep{WGT08}. This trio is evaluated in the order of decreasing flight duration. 

First, on examining the $LH_2$ \& $LOX$ option, it provides a useful improvement over solid propellant by reaching Planet 9 in 51 years utilizing a JOM (i.e., \emph{without} requiring a SOM), thereby exceeding the performance of the corresponding solid propellant JOM scenarios by at least 16\%. The SOM was not investigated for this propulsion option due to the close approach to the sun and the accompanying high solar flux, which would jointly be problematic for storage and deployment of cryogenic propellants.

Second, moving to NTP (again using the JOM option outlined above), the flight duration is noticeably reduced compared to solid chemical propellants, with an approximate 30\% reduction in flight time and around 20\% compared to the case of cryogenic chemical propulsion. The flight duration for this outcome is around 41 years. If the SOM option were to be feasible, the flight duration may be expected to decrease further.

Finally, along expected lines, laser sails have the superior performance by far, with a flight duration as low as 7 years. While this short timescale makes such mission designs an appealing option, the accompanying readiness level of laser sails is substantially lower than any of the alternative options investigated here. If the TRL of this propulsion scheme were to increase in the future, it is conceivable that laser-driven light sails would constitute the long-term future for the exploration of the outer Solar system and nearby interstellar space.

At this juncture, we will briefly examine the potential of other advanced (viz., non-chemical) propulsion systems, and how their performance may stack up against the mission architectures analyzed previously in this paper. We emphasize that this list is not exhaustive as it does not include, among others, nuclear fusion propulsion \citep{GK20,AGK21}, which might become viable in the future.
\begin{itemize}
\item Laser electric propulsion: Laser electric propulsion, as proposed by \citet{schmidt2018electric} and \citet{BPA18}, might enable velocities of $\sim 20$-$40$ AU/yr (entailing laser power of several 100 MW). If we assume such a terminal velocity for the bulk of the journey, this would permit fast trip times on the order of 10-20 years. Major technology development in connection with the laser array would, however, be required.
\item Solar sails: Solar sails, which operate on the same principle as laser sails except for using solar radiation as the power source instead of lasers, have been studied for several decades for deep space exploration \citep{McIn04}. Recent publications on the SunDiver mission concept of a kg-sized, low-cost solar sail mission estimate that, purely by using existing solar sail materials, hyperbolic excess velocities of up to 6 AU/yr could be achieved \citep{GFD22}. The mission duration would then be roughly the same as (or higher than) the architectures presented in this paper. However, the development of more sophisticated sail materials (see \citealt{ADI18}) could enable terminal velocities of $\sim 20$ AU/yr to be realized \citep{LL20}, which would then enable a mission to Planet 9 within approximately 23 years.\footnote{Even higher speeds for solar sails are attainable near high-energy astrophysical objects \citep{ML20}, but this scenario is obviously not applicable to the Solar system.}
\item Electric sails: Electric sails for deep space missions have previously been proposed in \citet{johnson2019electric}; this propulsion system was first introduced by \citep{Jan04}, and a modern review can be found in \citet{BNQ22}. The study by \citet{johnson2019electric} estimates that a 500 kg payload could be accelerated to a speed of 12 AU/yr via an electric sail; higher speeds of $\gtrsim 20$ AU/yr are feasible in theory \citep{JS07,LL20}. If this result is employed, this architecture would enable a mission to Planet 9 within $\sim 37$ years. However, scaling up an electric sail for such a mission constitutes an onerous challenge, given that in-orbit demonstration was only achieved for EstCube-1, a CubeSat-size spacecraft \citep{SPK15}.
\item Magnetic sails: Magnetic sails have been likewise proposed for deep space missions since their inception \citep{ZA91}, and are theoretically capable of attaining high velocities broadly comparable to electric sails \citep{LL21}. However, their performance has been subject to debates and is strongly dependent on the availability of robust advanced superconducting materials.
\end{itemize}
The consideration of advanced (non-chemical) propulsion systems studied hitherto in the paper shows that Planet 9 appears to be intriguingly poised at the transition point where chemical propulsion reaches its limit, indicating that advanced propulsion systems are rendered desirable and perhaps even necessary. 

Given that some of the technologies outlined hitherto such as sophisticated solar sails are potentially just 5-10 years away from development, it is plausible that even if we could launch a mission to Planet 9 today based on chemical propellant(s), such a spacecraft might be overtaken by a solar sail probe launched later. It would, therefore, represent the interplanetary analog of the waiting paradox, which is usually evoked for interstellar propulsion (see, e.g., \citealt{RH17}).

In closing, we reiterate that a scientific mission to characterize Planet 9 (should it prove to be real) has tremendous scientific value, with some of the chief benefits summarized in Section \ref{SciReturn}. If Planet 9 is indeed confirmed to be a planet with a mass between that of Earth and Neptune, even a flyby mission would yield a wealth of information regarding planet formation and dynamical evolution, the history of the Solar system, astrobiology, and much more. It is unlikely, in contrast, that these fields would experience advancements to the same degree if studies of Planet 9 were only restricted to data garnered from Earth- and space-based telescopes.

\section{Conclusion}
There has been renewed interest in the existence of Planet 9 and its basic properties, ever since the well-known work of \citet{BB16}. However, in light of its great distance from Earth (viz., semimajor axis of $\sim 400$ AU), a detailed characterization of this putative object may be difficult to accomplish from Earth. Thus, with this crucial limitation in mind, we explore a variety of mission architectures and trajectories to Planet 9, with the purpose of carrying out a flyby.

The various mission architectures invoked for reaching Planet 9 entail a combination of chemical propulsion (both solid propellant and liquid cryogenic propellant) and flyby maneuvers; the spacecraft payload specified in the paper is typically $\sim 100$ kg. The resulting mission duration for solid propellant ranges from 45 years to 75 years, depending on the distance from the Sun for the Solar Oberth maneuver; and for cryogenic propellant, a simple Jupiter Oberth maneuver would be sufficient to reduce transit times to approximately 50 years. These timescales are generally shorter than the flight times of 48 to 67 years obtained by \citet{CFP22} using a Jupiter gravity assist because the latter study utilizes a clearly optimistic semimajor axis of $300$ AU (see \citealt{BroB21}), whereas we adopt a conservative value of $450$ AU (as stated in Table \ref{table:Planet9}).

Looking beyond chemical propulsion, we also examine the prospects for reaching Planet 9 via more advanced, yet-to-be fully developed propulsion schemes. We find that nuclear thermal propulsion can reduce the mission duration to approximately 40 years. Further down the road, if the huge potential of laser sails is unlocked by humanity, rapid journey times of $\sim 7$ years are realizable. We also indicate how other near-future technologies such as laser electric propulsion, solar sails, and electric sails can enable flight times of $\sim 10$-$40$ years.

Thus, we are led toward the conclusion that Planet 9 comprises an object near the critical transition point where chemical propulsion approaches its performance limits (in the sense of supporting non-negative payloads) and alternative sophisticated propulsion systems (e.g., light sails) become seemingly more attractive vis-\`a-vis flight duration. Future work will necessitate analyzing the advanced propulsion schemes not investigated herein and explicating their mission architectures.

\acknowledgments

\bibliographystyle{aasjournal}
\bibliography{Planet9}

\end{document}